\PassOptionsToPackage{unicode}{hyperref}
\PassOptionsToPackage{hyphens}{url}
\PassOptionsToPackage{dvipsnames,svgnames,x11names}{xcolor}
\documentclass[
  12pt]{article}
  
\usepackage{multirow}
\usepackage{amsmath,amssymb}
\usepackage{iftex}
\ifPDFTeX
  \usepackage[T1]{fontenc}
  \usepackage[utf8]{inputenc}
  \usepackage{textcomp} 
\else 
  \usepackage{unicode-math}
  \defaultfontfeatures{Scale=MatchLowercase}
  \defaultfontfeatures[\rmfamily]{Ligatures=TeX,Scale=1}
\fi
\usepackage{lmodern}
\ifPDFTeX\else  
\fi
\IfFileExists{upquote.sty}{\usepackage{upquote}}{}
\IfFileExists{microtype.sty}{
  \usepackage[]{microtype}
  \UseMicrotypeSet[protrusion]{basicmath} 
}{}
\makeatletter
\@ifundefined{KOMAClassName}{
  \IfFileExists{parskip.sty}{%
    \usepackage{parskip}
  }{
    \setlength{\parindent}{0pt}
    \setlength{\parskip}{6pt plus 2pt minus 1pt}}
}{
  \KOMAoptions{parskip=half}}
\makeatother
\usepackage{xcolor}
\makeatletter
\ifx\paragraph\undefined\else
  \let\oldparagraph\paragraph
  \renewcommand{\paragraph}{
    \@ifstar
      \xxxParagraphStar
      \xxxParagraphNoStar
  }
  \newcommand{\xxxParagraphStar}[1]{\oldparagraph*{#1}\mbox{}}
  \newcommand{\xxxParagraphNoStar}[1]{\oldparagraph{#1}\mbox{}}
\fi
\ifx\subparagraph\undefined\else
  \let\oldsubparagraph\subparagraph
  \renewcommand{\subparagraph}{
    \@ifstar
      \xxxSubParagraphStar
      \xxxSubParagraphNoStar
  }
  \newcommand{\xxxSubParagraphStar}[1]{\oldsubparagraph*{#1}\mbox{}}
  \newcommand{\xxxSubParagraphNoStar}[1]{\oldsubparagraph{#1}\mbox{}}
\fi
\makeatother

\usepackage{longtable,booktabs,array}
\usepackage{calc} 
\usepackage{etoolbox}
\makeatletter
\patchcmd\longtable{\par}{\if@noskipsec\mbox{}\fi\par}{}{}
\makeatother
\IfFileExists{footnotehyper.sty}{\usepackage{footnotehyper}}{\usepackage{footnote}}
\makesavenoteenv{longtable}
\usepackage{graphicx}
\makeatletter
\def\maxwidth{\ifdim\Gin@nat@width>\linewidth\linewidth\else\Gin@nat@width\fi}
\def\maxheight{\ifdim\Gin@nat@height>\textheight\textheight\else\Gin@nat@height\fi}
\makeatother
\setkeys{Gin}{width=\maxwidth,height=\maxheight,keepaspectratio}
\makeatletter
\def\fps@figure{htbp}
\makeatother

\makeatletter
\@ifpackageloaded{caption}{}{\usepackage{caption}}
\AtBeginDocument{%
\ifdefined\contentsname
  \renewcommand*\contentsname{Table of contents}
\else
  \newcommand\contentsname{Table of contents}
\fi
\ifdefined\listfigurename
  \renewcommand*\listfigurename{List of Figures}
\else
  \newcommand\listfigurename{List of Figures}
\fi
\ifdefined\listtablename
  \renewcommand*\listtablename{List of Tables}
\else
  \newcommand\listtablename{List of Tables}
\fi
\ifdefined\figurename
  \renewcommand*\figurename{Figure}
\else
  \newcommand\figurename{Figure}
\fi
\ifdefined\tablename
  \renewcommand*\tablename{Table}
\else
  \newcommand\tablename{Table}
\fi
}
\@ifpackageloaded{float}{}{\usepackage{float}}
\floatstyle{ruled}
\@ifundefined{c@chapter}{\newfloat{codelisting}{h}{lop}}{\newfloat{codelisting}{h}{lop}[chapter]}
\floatname{codelisting}{Listing}

\makeatother
\makeatletter
\@ifpackageloaded{caption}{}{\usepackage{caption}}
\@ifpackageloaded{subcaption}{}{\usepackage{subcaption}}
\makeatother

\ifLuaTeX
  \usepackage{selnolig}  
\fi
\usepackage[authoryear,round]{natbib}
\usepackage{bookmark}

\IfFileExists{xurl.sty}{\usepackage{xurl}}{} 
\hypersetup{
  pdftitle={Title},
  pdfauthor={Author 1; Author 2},
  pdfkeywords={3 to 6 keywords, that do not appear in the title},
  colorlinks=true,
  linkcolor={blue},
  filecolor={Maroon},
  citecolor={Blue},
  urlcolor={Blue},
  pdfcreator={LaTeX via pandoc}}

\newcommand{\anon}{1}

\begin{document}

\def\spacingset#1{\renewcommand{\baselinestretch}%
{#1}\small\normalsize} \spacingset{1}


\ifnum\anon=1
{
  \title{\textbf{Spatial Analysis for AI-segmented Histopathology Images:\\ Methods and Implementation}}
  
  \author{
    Yoolkyu Park\textsuperscript{1},
    Fangjiang Wu\textsuperscript{2},
    Xin Feng\textsuperscript{1},
    Shengjie Yang\textsuperscript{2} \and
    Elizabeth H. Wang\textsuperscript{3},
    Bo Yao\textsuperscript{2},
    Chul Moon\textsuperscript{4},
    Guanghua Xiao\textsuperscript{2*}, and
    Qiwei Li\textsuperscript{1*}
  }
  
  \date{}
  
  \maketitle
  
  \noindent
  \begin{minipage}{\textwidth}
    \centering
    \small
    \textsuperscript{1}Department of Mathematical Sciences, The University of Texas at Dallas, Richardson, TX 75080  \\
    \textsuperscript{2}School of Public Health, The University of Texas Southwestern Medical Center, Dallas, TX 75390 \\
    \textsuperscript{3}Mays Business School, Texas A\&M University, College Station, TX 77843  \\
    \textsuperscript{4}Department of Statistics and Data Science, Southern Methodist University, Dallas, TX 75205 \\[0.5em]
    \textsuperscript{*}Corresponding authors
  \end{minipage}
  \vspace{1em}
  
}\fi

\if0\anon
{
  \bigskip
  \bigskip
  \bigskip
  \begin{center}
    {\LARGE\bf Title}
\end{center}
  \medskip
} \fi

\bigskip
\begin{abstract}
Quantitatively characterizing the spatial organization of cells and their interaction is essential for understanding cancer progression and immune response. Recent advances in machine intelligence have enabled large-scale segmentation and classification of cell nuclei from digitized histopathology slides, generating massive point pattern and marked point pattern datasets, which in turn create a growing need for accessible and standardized tools for downstream spatial statistical analysis.
However, 
accessible, reproducible, and integrative tools for exploratory and feature-based quantitative analysis
of such complex cellular spatial organization remain limited. In this paper, 
we do not aim to provide a comparative review or performance evaluation, but instead systematically organize and operationalize 27 widely used spatial summary statistics, areal indices, and topological features applicable to point pattern data within a unified computational framework.
We introduce SASHIMI (Spatial Analysis for Segmented Histopathology Images using Machine Intelligence), a browser-based tool for real-time spatial analysis of artificial intelligence (AI)-segmented histopathology images. SASHIMI enables the computation of a comprehensive suite of mathematically grounded descriptors, including spatial statistics, proximity-based measures, grid-level similarity indices, spatial autocorrelation measures, and topological descriptors, to quantify cellular abundance and cell-cell interaction. Applied to two cancer datasets, oral potentially malignant disorders (OPMD) and non-small-cell lung cancer (NSCLC), SASHIMI 
is used to illustrate the types of spatial and architectural patterns that can be captured, with several extracted features showing associations with patient survival outcomes.
SASHIMI provides an accessible and reproducible platform for single-cell-level spatial profiling of tumor morphological architecture, 
enabling interactive visualization and exploratory analysis of cellular spatial organization without requiring programming expertise.
\end{abstract}

\noindent%
\vfill

\newpage
\spacingset{1.8} 

\section{Introduction}\label{sec-intro}

Tissue morphology provides fundamental insights into organ function and disease mechanisms. In tumor tissues, the complex spatial arrangement of malignant and non-malignant cells forms the tumor microenvironment (TME)~\citep{hanahan2011hallmarks}, which directly influences cancer progression, immune response, and therapeutic outcomes~\citep{zhai2017spatial, wang2023deep}. Histopathology imaging remains the gold standard for visualizing cellular and tissue-level morphological architecture, offering rich histological detail on nuclei, lymphocytes, glands, and stromal structures, together providing ``a comprehensive view of disease and its effect on tissue”~\citep{gurcan2009histopathological}. Consequently, microscopic examination of hematoxylin and eosin (H\&E)-stained slides is an essential step in the diagnosis and staging of numerous diseases. Yet this workflow relies heavily on expert pathologists to detect and interpret subtle spatial and morphological patterns within highly complex images. The process is time-consuming, subjective, and prone to substantial inter and intra-observer variability~\citep{van2010interobserver, cooper2015novel}. 
As histopathology analysis increasingly shifts toward large-scale, cell-resolved digital workflows enabled by AI-based segmentation, there is a growing demand for quantitative and reproducible tools that enable a broader research community to characterize cellular spatial organization beyond expert-driven statistical programming environments.

The rapid advancement of artificial intelligence (AI) has reshaped histopathology imaging analysis, enabling quantitative, scalable, and reproducible assessment of complex tissue structures~\citep{SALTZ2018181, corredor2019spatialb, wang2019convpath, sobhani2022spatiala, wang2022spatialc}. 
Deep learning–based segmentation methods now delineate millions of individual cells from H\&E–stained slides and automatically predict phenotypic or morphological labels that previously required manual pathologist annotation~\citep{wang2019convpath}. 
Importantly, these advances primarily operate at the image segmentation and cell classification stage. Once segmentation is completed, downstream analysis of cellular spatial organization and cell–cell interactions relies on statistical and mathematical modeling of the resulting point pattern data rather than deep learning models themselves.
As a result, characterizing cellular spatial organization and cell–cell interactions has become a central objective in modern computational pathology, though it remains challenging due to the inherent complexity and heterogeneity of tissue architecture.
This shift highlights a critical methodological gap between AI-driven image segmentation and statistically principled spatial analysis that remains accessible to end users.


An AI-segmented histopathology image can be naturally represented as a marked point pattern, where each cell is characterized by its spatial coordinates and associated phenotypic or morphological label. This representation bridges digital pathology with spatial point pattern analysis and topological data analysis, providing a rigorous statistical framework for studying cellular spatial organization with the TME. 
A growing body of work shows that spatial correlation among cell types carries important prognostic value. For example, hierarchical Bayesian models have been used to quantify heterogeneous tumor–immune interactions, which are strongly associated with patient survival outcomes~\citep{li2019bayesiana, li2019bayesianb}. 
Additional distance-based, co-localization, and functional summary approaches have also been proposed to capture spatial heterogeneity and interaction patterns in multiplex and histopathology imaging data, including quantile-based biomarkers, spatial co-localization indices, and comparative evaluation frameworks \citep{Yi2023, Fridley2024, Soupir2025}.
More recent studies by \citet{moon2023using}, \citet{zhang2023deep}, \citet{stolz2024relational}, and \citet{zhang2024bayesian} demonstrate that geometric and topological descriptors capturing tissue shape can effectively stratify patient survival outcomes and characterize cellular and dynamical patterns. 
While these approaches focus on statistical modeling and inference, such as distance-matrix–based modeling, point process formulations, and functional ANOVA frameworks \citep{Masotti2023, Seal2024, Mohammed2024}, their practical application often requires extensive preprocessing, feature engineering, and custom implementations.
In parallel, AI-assisted frameworks have been developed to extract morphological and shape-based features from segmented histopathology images. For example, SAFARI~\citep{Fernandez2022SAFARI} focuses on quantifying nuclear shape characteristics from AI-segmented images to support downstream statistical analysis. While such approaches advance morphology-aware image analysis, they primarily operate at the level of shape representation and feature extraction, leaving the systematic characterization of cellular spatial organization and cell--cell interactions as a distinct and complementary challenge.
Taken together, these studies underscore the clinical importance of quantifying multiscale cellular and architectural organization in tumor tissues using complementary spatial and shape-based statistical descriptors.


Despite these methodological advances, there
remains a lack of integrative and accessible computational platforms
that enable researchers to efficiently quantify, visualize and explore cellular spatial organization across large-scale AI-segmented histopathology images. 
Rather than the absence of individual methods, the primary limitation lies in the fragmentation of existing spatial and topological techniques across disparate software packages and analytical pipelines, as also highlighted in recent statistical studies on quantitative cancer imaging \citep{Mohammed2024, Soupir2025}.
Recent efforts, such as spatial pathology features (SPF)~\citep{vu2022spf} and spatial quantification of pathology features (Spatial-QPFs)~\citep{li2024spatialqpfs}, have demonstrated the clinical value of spatial summary statistics. However, these approaches rely on a limited set of spatial summary statistics and are available primarily as offline \texttt{R} packages, 
which limits interactive exploration, standardized output, and rapid integration into downstream statistical workflows.

To address these limitations, 
we focus on the design and implementation of a unified software platform that organizes, standardizes, and operationalizes widely used spatial analytical techniques for histopathology imaging data. Rather than providing a comparative evaluation of methods, we incorporate $27$ established
spatial summary statistics, areal indices, and topological features applicable to point pattern data
into a single computational framework.
We introduce SASHIMI (Spatial Analysis of Segmented Histopathology Images using Machine Intelligence), a unified web-based platform that integrates these spatial and topological methods into a scalable, accessible, and reproducible analytical framework. SASHIMI focuses on integration, standardization, and real-time implementation of diverse analytical approaches for characterizing cellular spatial organization within tumor tissues. 
The platform encompasses three complementary analytical modules:
(i) spatial summary statistics, capturing intercellular proximity and interaction patterns;
(ii) topological and graph-based features, describing connectivity and structural complexity; and
(iii) areal similarity indices, quantifying spatial heterogeneity across tissue compartments.
Implemented as a real-time, browser-based interface, SASHIMI enables researchers to perform comprehensive exploratory spatial analyses without requiring extensive programming expertise.
Through two illustrative case studies: oral potentially malignant disorders (OPMD) from the Erlotinib Prevention of Oral Cancer (EPOC) trial and non–small-cell lung cancer (NSCLC) from the National Lung Screening Trial (NLST), we demonstrate how SASHIMI facilitates exploratory investigation of cellular spatial architecture and the generation of clinically relevant hypotheses. Together, these examples underscore SASHIMI’s role as a practical and reproducible platform for quantitative digital pathology.




The remainder of this paper is organized as follows.
Section~\ref{sec-data} describes (marked) point pattern data extracted from histopathology images.
Section~\ref{sec-features} introduces the imaging feature extraction framework, covering spatial summary statistics, areal data indices, and topological features.
Section~\ref{sec-tool} details the implementation of the SASHIMI online tool, including system design, input data requirements, output visualization modules, and user workflow.
Section~\ref{sec-case} presents two illustrative case studies demonstrating the use of SASHIMI for exploratory spatial analysis, followed by a sensitivity analysis examining how exploratory findings vary across commonly used significance thresholds.
Finally, Section~\ref{sec-conclusion} concludes the paper and discusses potential directions for future research.

\section{Data}\label{sec-data}

\subsection{Point Pattern Data}
Let $X = \{(x_{1}, y_{1}), \ldots, (x_{n}, y_{n})\}$ denote a spatial point pattern, where each coordinate
$(x_{i}, y_{i}) \in \mathbb{R}^{2}$ represents the coordinates of the $i$-th cell in a histopathology image, for $i = 1,\ldots, n$.

\subsection{Marked Point Pattern Data}
A marked point pattern extends the point pattern framework by assigning a cell type to each point. Let \(X = \{(x_{1}, y_{1}, m_{1}), \ldots, (x_{n}, y_{n}, m_{n})\}\) denote a marked point pattern, where \(m_{i} \in M\) represents the cell type of the $i$-th point, and \(M = \{\text{tumor}, \text{stromal}, \text{lymphocyte}, \ldots\}\) is the set of all observed cell types.

For each cell type \(m \in M\), we define \(X_m = \{(x_i, y_i)| m_i = m\}\) as the subset of cells with type $m$. These subsets form a partition of the full point pattern: \(X = \bigcup_{m\in M} X_m\). 
In the following, we use subscripts such as $X_{p}, X_{q}$ and $X_{r}$ to denote type-specific subsets, where \(p, q, r \in M\) indicate distinct cell types.

\section{Imaging Features}\label{sec-features}
We summarize $27$ scalar and functional descriptors below, covering spatial summary statistics, areal indices, and topological features commonly used in point pattern analysis. A concise overview of each descriptor, including its definition and interpretability, is provided in Supplementary Table 1.

\subsection{Spatial Summary Statistics}\label{sec-spatial-summary}

Spatial summary statistics are functions that analyze the spatial characteristics of points based on their locations and relationships. These functions characterize properties such as clustering, regularity, and inter-type spatial relationships~\citep{baddeley2016spatial}. 
Given marked point pattern data extracted from segmented pathology images, SASHIMI automatically computes the corresponding spatial summary functions and produces graphical visualizations.
Since computed spatial features are functional data, SASHIMI displays the computed functions alongside the input image, 
providing an intuitive visual interpretation of spatial organization.

The tumor tissue exhibits spatial heterogeneity across multiple scales, where 
cells may cluster locally while showing regular spacing at larger distances, and different cell types may exhibit attraction, repulsion, or independence. To capture this complexity, SASHIMI computes a comprehensive suite of spatial summary statistics organized into function families, providing both univariate analyses of single cell types and bivariate analyses of inter-type interactions. For example, Ripley's $K$-function~\citep{ripley1977modelling} 
quantifies the expected number of neighboring points within distance $r$ of a typical point, revealing clustering or dispersion patterns. The $G$-function characterizes the distribution of nearest-neighbor distances, while the $J$-function combines 
inter-point and empty-space information to detect deviations from complete spatial randomness.

All spatial summary statistics in SASHIMI are computed using the \texttt{R} package \texttt{spatstat}~\citep{baddeley2016spatial}, which implements rigorous edge-correction methods to mitigate boundary effects in finite observation windows. 

	\subsubsection{\textbf{$K$ and $L$-functions}} 
        The $K$-function is a second-order spatial summary statistic that quantifies the expected number of additional points within a distance \(r\) of a typical point, normalized by the overall intensity \(\lambda\) of the process. Formally, for a stationary point process, it is defined as:
        \[
            K(r) = \frac{1}{\lambda} \mathbb{E}[\text{number of further points within distance } r \text{ of a typical point}]
        \]
        Under a homogeneous Poisson process (i.e., complete spatial randomness), this simplifies to \(K(r) = \pi r^2\). 
        Deviations from this theoretical curve reveal departures from spatial randomness: when $K(r) > \pi r^{2}$, the pattern exhibits spatial clustering, whereas $K(r) < \pi r^{2}$ indicates spatial inhibition or regularity.

	\noindent \textbf{Ripley's $K$-function}
    
	Ripley's $K$-function is given by:
	\begin{equation}
		K(r) = \frac{1}{\lambda} \sum_{j}\sum_{i \neq j} I_r(d_{ij}) w_{i}, \quad \lambda = \frac{N}{A}
	\end{equation}
	where \( N \) is the total number of points and \( A \) is the area of the region, which is 1 when normalized. The Euclidean distance between two points \( (x_{i}, y_{i}) \) and \( (x_{j}, y_{j}) \) is \(d_{ij} = \sqrt{(x_{i} - x_{j})^2 + (y_{i} - y_{j})^2}\).
    The indicator function \( I_r(d_{ij}) \) equals 1 if \( d_{ij} \le r \) and 0 otherwise. Ripley’s isotropic correction weight $w_{ij}=1/e_{ij}$ provides an isotropic edge correction, where $e_{ij}$ denotes the proportion of the circumference of the circle centered at $(x_i, y_i)$ with radius equal to the inter-point distance $d_{ij}$ that lies within the observation window.

	\noindent \textbf{Directional $K$-function} 
    
	The directional $K$-function considers spatial patterns along specific directions:
	\begin{equation}
		K_{\theta}(r) = \frac{1}{\lambda}\sum_{j} \sum_{i \neq j} I_r(d_{ij}) \cdot I_{\theta}(\angle_{ij}) w_{ij}
	\end{equation}
	where \( I_{\theta}(\angle_{ij}) \) ensures that only pairs within the directional range \( \theta - \frac{\Delta          \theta}{2} \leq \angle_{ij} < \theta + \frac{\Delta \theta}{2} \) are considered.\\

	\noindent \textbf{Cross-type $K$-function}
    
	The Cross-type $K$-function evaluates spatial relationships between different categories of points:
	\begin{equation}
        K_{\text{cross}}(r) = \frac{1}{\lambda_q}\sum_{i\in X_p}\sum_{j\in X_q} I_r(d_{ij})w_{ij}
	\end{equation}
	where \( \lambda_q \) denotes the intensity of the point process of type \( q \).

	\noindent \textbf{Mark Weighted $K$-function} 
    
	The mark weighted $K$-function extends Ripley's $K$-function by incorporating weights based on spatial point attributes, its mark value:
    \begin{equation}
    K_f(r) = \frac{1}{\lambda\, f(M,M')} 
    \sum_{i}\sum_{j\ne i} f(m_i,m_j)\, I_r(d_{ij}),
    \end{equation}
    where $f(m_i,m_j)$ is a user-defined weight function (by default $f(m_i,m_j)=m_i m_j$).

	\noindent \textbf{Scaled $K$-function ($L$-function)}
    
	The scaled $K$-function, also known as the $L$-function, normalizes $K(r)$ to remove dimensional effects:
	\begin{equation}
		L(r) = \sqrt{\frac{K(r)}{\pi}}
	\end{equation}
    so that $L(r)>r$ indicates spatial clustering and $L(r)<r$ indicates spatial inhibition or regularity. 
	This function normalizes the traditional $K$-function, removing dimensionality effects.

	\noindent \textbf{Cross-type $L$-function}
    
	The Cross-type $L$-function captures spatial correlations between points with different marks, extending the concept of      the scaled $K$-function:
	\begin{equation}
		L_{\text{cross}}(r) = \sqrt{\frac{K_{\text{cross}}(r)}{\pi}}
	\end{equation}

	\subsubsection{\textbf{$G$, $F$, and $J$-Functions}} 
    
	
    The $G$-function, or nearest-neighbor distribution function, describes the cumulative distribution of distances from each point in the pattern to its nearest neighboring point. It takes values in $[0,1]$. Under complete spatial randomness (CSR) with intensity~$\lambda$, the theoretical form is 
    \begin{equation}
    G_{\text{CSR}}(r) = 1 - \exp(-\lambda \pi r^2).
    \end{equation}
    Departures from this theoretical curve reveal deviations from spatial randomness: when $G(r)$ lies above $G_{\text{CSR}}(r)$, the pattern exhibits clustering (an excess of short inter-point distances), whereas values below $G_{\text{CSR}}(r)$ indicate inhibition (a deficiency of close neighbors).
    Additionally, $F$-function follows the same nature. 
    
	\noindent \textbf{$G$-function}

	The $G$-function captures empty spaces between points by measuring the nearest neighbor distance from each point. 
    The empirical estimator of $G(r)$ is
    \begin{equation}
    G(r) = \frac{1}{N}\sum_{i=1}^{N} I_r(d_i),
    \end{equation}
    where $d_i = \min_{j \ne i} d_{ij}$ is the distance from point $i$ to its nearest neighbor, $d_{ij} = \sqrt{(x_i-x_j)^2+(y_i-y_j)^2}$ is the Euclidean distance between points $i$ and $j$, and $I_r(d_i)=1$ if $d_i\le r$ and $0$ otherwise.

	\noindent \textbf{Cross-type $G$-Function}
    
    The Cross-type $G$-function quantifies the distribution of nearest-neighbor distances from points of type $p$ to points of type $q$:
    \begin{equation}
      G_{\text{cross}}(r) = \frac{1}{N_p} \sum_{i \in X_p} I_r(d_{i,q}),
    \end{equation}
    where for each point $i \in X_p$, the distance $d_{i,q}$ is defined as $d_{i,q} = \min_{j \in X_q} d_{ij}$. Here, $I_r(d_{i,q}) = 1$ if $d_{i,q} \le r$ and $0$ otherwise.

	\noindent \textbf{$F$-function}
    
	The $F$-function, similarly, measures the empty space in the point process pattern, but uses the distance from a randomly chosen point \( u \) to its nearest neighbor:
	\begin{equation}
		F(r) = \frac{1}{N}\sum_{u} I_{r}(d(u))
	\end{equation}
	where the empty-space distance from a randomly chosen reference point \( u \) to its nearest neighbor \( i \) is $d(u) = \min\{\lVert (x_{u},y_{u}) - (x_{i},y_{i}) \rVert\}$.
    
	\noindent \textbf{Cross-type $F$-function}
    
	The Cross-type $F$-function measures the distance from a randomly reference chosen point \( u \) to its nearest type \( q \) neighbor \( j \):
	\begin{equation}
		F_{\text{cross}}(r) = \frac{1}{N}\sum_{u} I_{r}(d(u, j))
	\end{equation}
	where the empty-space distance from a randomly chosen reference point \( u \) to its nearest type \( q \) neighbor \( j \) is $d(u ,j) = \min\{\lVert (x_{u},y_{u}) - (x_{j},y_{j}) \rVert\}$.

	\noindent \textbf{$J$-function}
    
	The $J$-function is a dimensionless ratio of two functions, \( G(r) \) and \( F(r) \), and is given by:
	\begin{equation}
		J(r) = \frac{1 - G(r)}{1 - F(r)}
	\end{equation}
    Under CSR, $J(r) \equiv 1$ for all $r$. 
    Values of $J(r)$ below 1 indicate clustering (an excess of short distances), 
    whereas values above 1 indicate regularity or inhibition. 
    Because it integrates both the inter-point ($G$) and empty-space ($F$) perspectives, the $J$-function is less sensitive to variations in point density 
    and provides a scale-independent measure of spatial structure.
    
	\noindent \textbf{Cross-type $J$-function}
    
	Similarly, the Cross-type $J$-function is given by:
	\begin{equation}
		J_{\text{cross}}(r) = \frac{1 - G_{\text{cross}}(r)}{1 - F_{\text{cross}}(r)}
	\end{equation}

	\subsubsection{\textbf{Correlation Functions}}
	
	\noindent \textbf{Pair Correlation Function (PCF)}
    
	The pair correlation function (PCF) provides a noncumulative measure of spatial dependence, complementary to Ripley’s $K$-function. While $K(r)$ summarizes the cumulative number of neighboring points within distance~$r$, the PCF, denoted $\text{pcf}(r)$, captures the local intensity of points exactly at distance $r$ from a typical point. Formally, it is defined as
    \begin{equation}
        \text{pcf}(r) = \frac{1}{2\pi r}\,\frac{dK(r)}{dr},
    \end{equation}
    where $K(r)$ is the Ripley’s $K$-function introduced previously.
    For a homogeneous Poisson process under complete spatial randomness (CSR), $\text{pcf}(r)\equiv1$. 
    Values $\text{pcf}(r)>1$ indicate spatial clustering (an excess of pairs at distance $r$), whereas $\text{pcf}(r)<1$ indicate inhibition or regularity (a deficit of pairs at that scale).

	\noindent \textbf{Cross-type Pair Correlation Function} 
    
	The Cross-type PCF, denoted by $\text{pcf}_{\text{cross}}(r)$, quantifies the spatial dependence between two distinct point types. It is obtained analogously from the Cross-type $K$-function:
	\begin{equation}
    \text{pcf}_{\text{cross}}(r)
      = \frac{1}{2\pi r}\,\frac{dK_{\text{cross}}(r)}{dr}.
    \end{equation}
    Here, $K_{\text{cross}}(r)$ measures the cumulative spatial interaction between two cell types, and its derivative $\text{pcf}_{\text{cross}}(r)$ reflects the strength of inter-type correlation at distance $r$. Under CSR, $\text{pcf}_{\text{cross}}(r)=1$ for all $r$; values above one suggest attraction between the two types, whereas values below one imply spatial segregation.
    
	\noindent \textbf{Mark Connected Function}
    
    When each point carries a quantitative or categorical mark $m_i$, the mark correlation function extends the PCF to assess correlations among marks at given distances. Let $f(m_i,m_j)$ denote a user-defined weighting function (e.g., product, indicator, or correlation of marks). The MCF is then expressed as
	\begin{equation}
		\text{mcf}_{r} = \frac{\lambda_{p}\lambda_{q}\text{pcf}_{cross}(r)}{\lambda^2 \text{pcf}(r)}
	\end{equation}
    where $\lambda_p$ and $\lambda_q$ are the intensities of the corresponding point types, $\lambda$ is the overall process intensity. Values $\text{mcf}_{r}>1$ indicate positive mark association (similar marks occur more frequently nearby), whereas $\text{mcf}_{r}<1$ suggest negative association or mark repulsion.

\subsection{Areal Data Indices}\label{sec-areal}

\addtolength{\textheight}{.2in}%

SASHIMI also computes spatial autocorrelation and similarity indices based on grid-level 
representations of the tissue. The marked point pattern data are tessellated into a $Q\times Q$ quadrat count matrix (e.g., $Q=20$), transforming the continuous spatial distribution into a discrete 
areal format. This representation enables quantification of spatial structure through 
statistics that assess cell-type aggregation, dispersion, and compositional similarity 
across tissue regions.

We denote by \(Q_X\) a $20 \times 20$ quadrat count grid computed from the original point pattern $X$.
    Type-specific quadrat counts are written as \(Q_{Xp}, Q_{Xq}, Q_{Xr}\), etc.
    Let \(Q_{X}[i]\) denote the count in the \(i\)th quadrat of \(Q_{X}\), and let \(n\) be the number of quadrats (\(n=400\) for a \(20\times20\) grid).

\subsubsection{Spatial Autocorrelation Measures} Spatial autocorrelation measures quantify the degree of spatial dependence 
between neighboring quadrats. These include Moran's I~\citep{moran1950notes}, Geary's C~\citep{geary1954contiguity}, and the Quadrat Count Statistic, which assess whether 
quadrats with similar cell densities tend to cluster spatially or exhibit random or dispersed patterns.

\noindent \textbf{Moran's I}
    
    Moran's I is another widely used measure of global spatial autocorrelation and is inversely related to Geary’s C. It is defined as
	\begin{equation}
		\text{MoranI} = \frac{N}{S_{0}}\frac{ \sum_i \sum_j w_{ij} (x_{i} - \bar{x})(x_{j} - \bar{x})}{\sum_i (x_{i} - \bar{x})^2}
	\end{equation}
	where the notations \(x_i\), \(\bar{x}\), and \(w_{ij}\) follow those in the definition of Geary’s C. Positive values of Moran’s I indicate clustering of similar cell densities, values near 0 indicate randomness, and negative values suggest spatial dispersion.
    
\noindent \textbf{Geary's C}
    
	Geary's C is a global measure of spatial autocorrelation that evaluates the similarity between neighboring quadrats. 

    Let \(N\) denote the total number of quadrats, \(x_i\) the cell count in the \(i\)th quadrat of \(Q_X\), and \(\bar{x}\) the mean of \(x_i\). Then Geary’s C is defined as
	\begin{equation}
		\text{GearyC} = \frac{(N-1)\sum_{i}\sum_{j}w_{ij}(x_{i}-x_{j})^2}{2S_{0}\sum_{i}(x_{i} - \bar{x})^2}
	\end{equation}
 where the spatial weights matrix is defined as \(w_{ij} = 1\) if quadrat \(i\) is a neighbor of \(j\), and \(w_{ij} = 0\) otherwise. The total sum of weights is \(S_0 = \sum_{i=1}^{N}\sum_{j=1}^{N} w_{ij}\).
        Values of Geary’s C range from 0 to values greater than 1. 
        A value close to 0 indicates strong positive spatial autocorrelation (neighboring quadrats are similar), while values greater than 1 indicate negative spatial autocorrelation (neighboring quadrats are dissimilar).

    \noindent \textbf{Lee’s \(L\)}
        
	Lee's \(L\) is a spatial correlation coefficient that measures the association between two sets of observations in the same spatial domain.
    Unlike the standard Pearson correlation, Lee’s \(L\) incorporates spatial dependence \textit{via} a spatial weights matrix. Let \(x_i\) and \(y_i\) denote the counts of two different cell types in the \(i\)th quadrat, and \(\bar{x}\), \(\bar{y}\) their respective means. Lee’s \(L\) is given by
	\begin{equation}
		\text{Lee'sL} =
		\frac{N}{\sum_i \left( \sum_{j} w_{ij} \right)^2}
		\frac{\sum_{i}\sum_{j} (\tilde{x}_{i} - \bar{x}) (\tilde{y}_{i} - \bar{y})}
		{\sqrt{\sum_{i} (\tilde{x}_{i} - \bar{x})^2} \sqrt{\sum_{i} (\tilde{y}_{i} - \bar{y})^2}}
	\end{equation}
	where $\tilde{x}_{i} = \sum_{j}w_{ij}x_{j}$, $\tilde{y}_{i} = \sum_{j}w_{ij}y_{j}$,
    and \(w_{ij}\) denotes spatial adjacency between quadrats \(i\) and \(j\). A higher Lee’s \(L\) value indicates stronger positive spatial correlation between the two variables.

    \noindent \textbf{Quadrat Count Statistic}
    
    The quadrat test assesses deviations from complete spatial randomness (CSR) by comparing observed and expected cell counts across quadrats:
	\begin{equation}
		X^2 = \sum_{i = 1}^{n} \frac{(O_{i}-E_{i})^2}{E_{i}}
	\end{equation}
	where \(O_{i}\) and \(E_{i}\) are the observed and expected counts in the \(i\)th quadrat under CSR. The test statistic \(X^{2}\) is compared to a chi-squared reference distribution to evaluate the null hypothesis \(H_{0}\): quadrat counts follow a Poisson distribution implied by CSR.


	\noindent \textbf{Join Count Statistic}
    
	The join count statistic evaluates spatial association (autocorrelation) for binary categorical maps. For a chosen pair of cell types \(p\) and \(q\), define a binary indicator \(b_{i}\in\{0,1\}\) on each quadrat (e.g., \(b_{i}=1\) if type \(p\) is present in the \(i\)th quadrat and \(b_{i}=0\) otherwise), and let \(w_{ij}\) be the spatial adjacency indicator between quadrats \(i\) and \(j\):
    \begin{equation}
        w_{ij} =
        \begin{cases}
            1, & \text{if } i \text{ is a neighbor of } j,\\
            0, & \text{otherwise.}
        \end{cases}
    \end{equation}
	The counts of like–like and unlike joins are
    \begin{equation}
        J_{pp} = \frac{1}{2}\sum_{i\neq j} w_{ij}\, b_{i} b_{j}, \qquad
        J_{pq} = \frac{1}{2}\sum_{i\neq j} w_{ij}\,(b_{i}-b_{j})^{2}, \qquad
        J_{qq} = \frac{1}{2}\sum_{i\neq j} w_{ij}\,(1-b_{i})(1-b_{j}),
    \end{equation}
    and the total number of joins is \(J = J_{pp}+J_{pq}+J_{qq}\). Here the factor \(1/2\) avoids double counting under an undirected neighbor graph. The choice of contiguity (e.g., rook vs.\ queen) determines \(w_{ij}\).

\subsubsection{Similarity Indices} 
Similarity indices measure compositional overlap or dissimilarity between spatial distributions of different cell types. SASHIMI computes the Jaccard Index, the Morisita-Horn Index~\citep{morisita1962sigma}, and cosine similarity to quantify the degree of co-localization among cell types 
within tissue regions.

\noindent \textbf{Tanimoto Coefficient}
    
	The Tanimoto coefficient (generalized Jaccard for vectors) index measures the similarity between type-specific quadrat count vectors and is computed only on quadrats where at least one of the two types is present. 
    For types \(p\) and \(q\), let \(Q_{Xp}\) and \(Q_{Xq}\) be the corresponding quadrat count vectors (restricted to the selected quadrats). The coefficient is
	\begin{equation}
        T(Q_{Xp}, Q_{Xq}) = \frac{Q_{Xp}^\top Q_{Xq}}
        {\|Q_{Xp}\|_2^2 + \|Q_{Xq}\|_2^2 - Q_{Xp}^\top Q_{Xq}}
        \end{equation}
	with range
    $0 \le T(Q_{Xp}, Q_{Xq}) \le 1$, where \(T=1\) indicates identical distributions and \(T=0\) indicates no overlap.
	
	\noindent \textbf{Dice - Sorensen Index (Vector Form)}
    
	The Sorensen index quantifies similarity between two quadrat count distributions and can be expressed in vector form as:
	\begin{equation}
        \text{DSC}(Q_{Xp}, Q_{Xq}) = \frac{2 Q_{Xp}^\top Q_{Xq}}
        {\|Q_{Xp}\|_2^2 + \|Q_{Xq}\|_2^2}
        \end{equation}    
        When the counts are binary, this expression reduces to the classical Dice–Sorensen index. The coefficient ranges between 0 and 1, where 1 indicates identical distributions and 0 indicates no overlap.

	\noindent \textbf{Morisita-Horn Index}
    
        The Morisita–Horn index measures compositional overlap between two quadrat count distributions, accounting for differences in abundance:
	\begin{equation}
		\text{MH} = \frac{2\sum_{i=1}^{N}Q_{Xp}[i] Q_{Xq}[i] }{\sum_{i=1}^N (Q_{Xp}[i])^2+\sum_{i=1}^N (Q_{Xq}[i])^2}
	\end{equation}
    Prior to computation, both quadrat count vectors are normalized by their total counts:

        where both quadrats \(Q_{Xp}, Q_{Xq}\) are normalized as:
        \begin{equation}
            Q_{Xp} = \frac{Q_{Xp}}{|Q_{Xp}|}, \qquad
            Q_{Xq} = \frac{Q_{Xq}}{|Q_{Xq}|}.
        \end{equation}
        The index takes values in \([0,1]\), where higher values indicate greater similarity between spatial compositions.

	\noindent \textbf{Clark and Evans Aggregation Index} 
    
	The Clark and Evans Aggregation Index quantifies the degree of clustering or regularity in a spatial point process:
	\begin{equation}
		\text{R} = \frac{\bar{r}_{o}}{\bar{r}_{e}}
	\end{equation}
	where \( \bar{r}_{o} \) is the average of the observed nearest distance from point \( i \) and 
\(\bar{r}_{e}\) is its theoretical expectation under complete spatial randomness (CSR). The observed mean distance is computed as $\bar{r}_{o} = \frac{1}{N}\sum d_{i}$, where \( N \) is the total number of points and \( d_{i} \) is the nearest-neighbor distance from point \( i \). Under CSR, the expected mean nearest-neighbor distance is $\bar{r}_{e} = \frac{1}{2 \sqrt{\lambda}}$ where \(\lambda\) is the intensity of the point pattern. An \(R\) value close to 1 indicates a random (CSR) distribution; 
\(R<1\) suggests clustering or aggregation; and \(R>1\) indicates spatial regularity or inhibition.


    \noindent \textbf{Bhattacharyya Coefficient}
    
	The Bhattacharyya coefficient 
    quantifies the degree of overlap between two discrete probability distributions. For two normalized quadrat count distributions \(Q_{Xp}\) and \(Q_{Xq}\), it is defined as
	\begin{equation}
		\text{BC} = \sum_{i} \sqrt{Q_{Xp}[i]Q_{Xq}[i]}
	\end{equation}
	where:
	\begin{equation*}
		Q_{Xp}[i] = \frac{\text{number of points in the } i\text{-th grid in \(Q_{Xp}\) }}{\text{number of all points in \(Q_{Xp}\) }}
	\end{equation*}
	\begin{equation*}
		Q_{Xq}[i] = \frac{\text{number of points in the } i\text{-th grid in \(Q_{Xq}\) }}{\text{number of all points in \(Q_{Xq}\) }}
	\end{equation*}
    The coefficient ranges from $0$ to $1$, where \(\text{BC} = 1\) indicates identical distributions and \(\text{BC} = 0\) indicates no overlap.

    \noindent \textbf{Cosine Similarity} 

        Cosine similarity measures the angular similarity between two nonzero vectors. For two feature vectors \(Q_{Xp}\) and \(Q_{Xq}\), it is defined as
        \begin{equation}
            \text{Cossim}(Q_{Xp}, Q_{Xq}) = \frac{\sum_{i} Q_{Xp}[i] Q_{Xq}[i]}{\sqrt{\sum_{i} (Q_{Xp}[i])^2} \sqrt{\sum_{i} (Q_{Xq}[i])^2}}.
        \end{equation}
        Its values range from $-1$ to $1$, where $1$ indicates perfect alignment (maximum similarity), $0$ indicates orthogonality (no relationship), and $-1$ indicates perfect opposition.

Together, these measures provide complementary perspectives on spatial organization: 
autocorrelation statistics capture global patterns of spatial dependence, while similarity indices describe local compositional relationships. All computed metrics are returned as scalar features in a single-row data frame. Computations are supported by the \texttt{R} packages \texttt{spatstat}~\citep{baddeley2016spatial} and \texttt{spdep}~\citep{bivand2015spdep}.

\addtolength{\textheight}{-.2in}%

\subsection{Topological Features}\label{sec-topology}
Topology, a branch of mathematics that focuses on spaces that are invariant under continuous transformations, forms the basis of topological data analysis (TDA). It enables characterization of spatial patterns through topological features such as the number of connected components (dimension zero features) and loops (dimension one features). In SASHIMI, topological features of marked point pattern data are computed using persistent homology with the witness complex, a method that approximates the shape of the data by encoding its local connectivity structure~\citep{de2004topological}. The resulting output is represented as a multiset of birth–death intervals. From these intervals, we extract summary statistics that quantitatively characterize the topological features across multiple spatial scales. For each dimension, we compute distributional statistics of birth times, death times, and feature lifetimes, including their minima, maxima, means, and standard deviations, along with the total number of detected features. This functional representation provides a complementary view of the spatial organization beyond classical distance-based descriptors. Computations are performed using the Python package \texttt{GUDHI} \citep{gudhi:urm}.


    \noindent \textbf{Witness Complex} 

    Let $L \subset X$ denote a set of \textit{landmark points} and $W \subset X$ a set of \textit{witness points}, where $X$ is the full point pattern. 
    The witness complex \(W(L, W, \varepsilon)\) approximates the geometric shape of the data by encoding its local connectivity structure.
    Formally, a $k$-dimensional simplex \(\sigma = \{l_0, \dots, l_k\} \subseteq L\) is included in the witness complex if there exists a witness \(w \in W\) such that all vertices of \(\sigma\) lie within a distance \(\varepsilon\) of \(w\):
    \[
    \exists\, w \in W \quad \text{s.t.} \quad 
    \max_{l_i \in \sigma} d(w, l_i) \le \varepsilon.
    \]
    This construction efficiently captures the topological relationships between the landmark and witness subsets while reducing computational complexity relative to full Vietoris–Rips complexes.

    In our application, cells of type A are selected as landmarks and cells of type B as witnesses. This asymmetric construction emphasizes how one cell population is geometrically arranged relative to the other. We use three types of cells for the witness complex: tumor, immune, and stromal. Therefore, the resulting persistent homology reflects cross-scale structural interplay, such as between tumor and immune or stromal components. \\

\subsection{Model-based Spatial Analysis Methods}

In addition to summary-based spatial descriptors, model-based approaches have been developed to analyze cellular spatial organization in histopathology and multiplex imaging data. These methods specify probabilistic or functional models under explicit structural assumptions, enabling formal statistical inference \citep{Masotti2023, Seal2024, Mohammed2024}.

Such approaches typically require model specification (e.g., parametric interaction functions or prior distributions), likelihood- or Bayesian-based estimation procedures, and tailored preprocessing steps. To contextualize the broader methodological landscape, several representative formulations are summarized below. These methods are not implemented within SASHIMI, which focuses on standardized spatial descriptors designed for scalable and interactive exploratory analysis.

\subsubsection{Bayesian Hidden Potts Mixture Model}

The Bayesian hidden Potts mixture model captures cellular heterogeneity and spatial dependence through a finite mixture model with Potts-type spatial regularization. Each spatial location $(x_i, y_i)$ is associated with a latent class label $m_i \in \{1,\dots,K\}$. Conditional on class membership, cell-level feature measurements are modeled via a finite mixture likelihood
\begin{equation*}
p(\boldsymbol{z} \mid \boldsymbol{m}, \boldsymbol{\theta})
=
\prod_{i=1}^{n}
f(z_i \mid \theta_{m_i}),
\end{equation*}
where $z_i$ denotes observed features at location $i$, $\theta_{m_i}$ are class-specific parameters, and $\boldsymbol{\theta}=\{\theta_k\}_{k=1}^K$.
Spatial dependence among neighboring latent labels is modeled using a Potts prior
\begin{equation*}
p(\boldsymbol{m} \mid \beta)
\propto
\exp\!\left(
\beta
\sum_{i \sim j}
\mathbb{I}(m_i = m_j)
\right),
\end{equation*}
where $i \sim j$ denotes neighboring locations and $\beta$ controls the strength of spatial smoothing. Bayesian inference is conducted under the joint posterior
\begin{equation*}
p(\boldsymbol{m}, \boldsymbol{\theta}, \beta \mid \boldsymbol{z})
\propto
p(\boldsymbol{z} \mid \boldsymbol{m}, \boldsymbol{\theta})
\,
p(\boldsymbol{m} \mid \beta)
\,
p(\boldsymbol{\theta})
\,
p(\beta),
\end{equation*}
allowing estimation of spatially structured latent classes and interaction strength \citep{Li2019a}.

\subsubsection{Bayesian Mark Interaction Model}

A Bayesian mark interaction model specifies a Gibbs distribution for the mark configuration conditional on observed spatial locations. Given the marked point pattern $\{(x_i,y_i,m_i)\}_{i=1}^n$, the marks are treated as random variables conditional on fixed locations. The model is defined through a local energy function
\begin{equation*}
V(\boldsymbol{m} \mid \boldsymbol{\omega}, \boldsymbol{\Theta}, \lambda)
=
\sum_{q}
\omega_q
\sum_{i}
\mathbb{I}(m_i = q)
+
\sum_{q,q'}
\theta_{qq'}
\sum_{i \sim j}
e^{-\lambda d_{ij}}
\mathbb{I}(m_i=q, m_j=q'),
\end{equation*}
where $\boldsymbol{\omega}=\{\omega_q\}$ are first-order intensity parameters, $\boldsymbol{\Theta}=\{\theta_{qq'}\}$ are second-order interaction parameters indexed by cell-type pairs, and $\lambda$ controls exponential decay of interaction strength with distance $d_{ij}$. The corresponding likelihood takes the Gibbs form
\begin{equation*}
p(\boldsymbol{m} \mid \boldsymbol{\omega}, \boldsymbol{\Theta}, \lambda)
=
\frac{
\exp\!\left[-V(\boldsymbol{m} \mid \boldsymbol{\omega}, \boldsymbol{\Theta}, \lambda)\right]
}{
Z(\boldsymbol{\omega}, \boldsymbol{\Theta}, \lambda)
},
\end{equation*}
where $Z(\cdot)$ is an intractable normalizing constant. Under a Bayesian framework, prior distributions are specified for $\boldsymbol{\omega}$, $\boldsymbol{\Theta}$, and $\lambda$, and inference is performed on their posterior distribution \citep{Li2019b}.

\subsubsection{DIMPLE: Distance-matrix–based Modeling Approaches}

DIMPLE summarizes inter-type spatial relationships through empirical distance distributions. For each pair of cell types $(p,q)$, inter-point distances $d_{ij}$ with $(x_i,y_i)\in X_p$ and $(x_j,y_j)\in X_q$ are computed and aggregated into a distribution $F_{pq}(r)$.

For two samples $a$ and $b$, differences between their interaction patterns are quantified using information-theoretic divergences. The Kullback–Leibler divergence is defined as
\begin{equation*}
\mathrm{KLD}\!\left(F_{pq}^{(a)} \,\|\, F_{pq}^{(b)}\right)
=
\sum_r 
F_{pq}^{(a)}(r)
\log 
\frac{F_{pq}^{(a)}(r)}
     {F_{pq}^{(b)}(r)},
\end{equation*}
and the symmetric Jensen–Shannon distance is
\begin{equation*}
\mathrm{JSD}\!\left(F_{pq}^{(a)}, F_{pq}^{(b)}\right)
=
\sqrt{
\frac{1}{2}
\mathrm{KLD}\!\left(F_{pq}^{(a)} \,\|\, M\right)
+
\frac{1}{2}
\mathrm{KLD}\!\left(F_{pq}^{(b)} \,\|\, M\right)
},
\quad 
M=\frac{1}{2}\!\left(F_{pq}^{(a)}+F_{pq}^{(b)}\right).
\end{equation*}
For multiple samples, these divergences are assembled into sample-by-sample distance matrices for each interaction type $(p,q)$, and statistical inference is performed on these matrices to assess associations with group labels or clinical outcomes \citep{Masotti2023}.

\subsubsection{SpaceANOVA: Point Process and Functional Modeling Approaches}

SpaceANOVA models inter-type spatial dependence through distance-dependent co-occurrence functions combined with functional ANOVA. For two cell types $p$ and $q$, a co-occurrence function $C_{pq}(r)$ summarizes interaction intensity between $X_p$ and $X_q$ at spatial scale $r$, estimated from inter-point distances $d_{ij}$. For multiple samples indexed by $k$, the co-occurrence function is treated as a functional response:
\begin{equation*}
C_{pq}^{(k)}(r)
=
\mu_{pq}(r)
+
\alpha_{pq,g(k)}(r)
+
\varepsilon_k(r),
\end{equation*}
where $\mu_{pq}(r)$ is a baseline interaction function, $\alpha_{pq,g(k)}(r)$ denotes group-specific functional effects, and $\varepsilon_k(r)$ is a stochastic error process. Functional ANOVA is used to test for differences in spatial interaction patterns across groups while accounting for spatial scale \citep{Seal2024}.


\section{Online Tool Implementation}\label{sec-tool}

\subsection{Implementation}
SASHIMI is a browser-based computational framework implemented in \texttt{R} and \texttt{Python} for the extraction, visualization, and computation of spatial features from AI-segmented histopathology images, enabling real-time analysis.
The web-based design is intentionally practitioner-facing: analyses are executed through the browser interface without requiring end-users to install a programming compiler, manage external package dependencies, or configure local environments. This deployment choice directly addresses common barriers for translational users who wish to interrogate AI-segmented pathology outputs but lack specialized computational infrastructure. Consistent with prior work on practitioner-oriented post-segmentation analytics~\citep{Fernandez2022SAFARI}, SASHIMI treats AI-segmented results as the analysis-ready input and focuses on producing interpretable spatial and topological descriptors that can be used immediately for exploratory investigation or downstream modeling.
The platform functions as both an exploratory and feature-extraction tool, generating two complementary forms of output from marked point pattern data:
(i) functional visualizations, which characterize distance-dependent spatial dynamics (e.g., Ripley’s $K$-function and pair correlation function), and
(ii) scalar indices, which quantify spatial autocorrelation and compositional similarity across tissue regions (e.g., Moran’s I, Jaccard index, and summary statistics derived from persistence diagrams).
Together, these outputs provide a comprehensive and interpretable representation of the spatial organization.

\subsection{Input Requirements and Data Format}
SASHIMI accepts tabular data in CSV format with a maximum file size of $4$ MB. 
Inputs are assumed to be derived from upstream AI-based segmentation pipelines (i.e., preprocessed cell detections and type annotations), so SASHIMI does not operate on raw whole-slide images; rather, it consumes segmentation-derived cell-level coordinates and labels as a standardized interchange format for post-segmentation spatial analysis.
Each input file must contain three columns corresponding to the $x$-coordinate, $y$-coordinate, and cell type. Upon upload, the data are displayed as a point cloud, where each point represents an individual cell on the tissue slide and is color-coded by cell type. Users may select up to three cell types for analysis. For each selection, SASHIMI computes both single-type spatial features and pairwise Cross-type spatial functions.

\subsection{Output Types and Visualization}
SASHIMI produces two forms of output. Functional outputs consist of spatial summary statistics rendered as plots, enabling interpretation of distance-based spatial relationships. 
Scalar outputs include autocorrelation measures, similarity indices, and summary statistics of the persistence diagram. Returned in tabular form, it provides quantitative measures suitable for direct comparison and statistical analysis. All computed features can be downloaded in CSV format for downstream use. 
By coupling immediate visualization with downloadable, analysis-ready feature tables, SASHIMI is designed to support rapid exploratory workflows for practitioners—allowing users to screen spatial patterns, generate hypotheses, and export standardized outputs for subsequent statistical modeling without additional software setup on the user side.

\subsection{User Workflow}

\begin{figure}[htbp]
\centering
\makebox[\textwidth][c]{
  \includegraphics[width=1.2\textwidth]{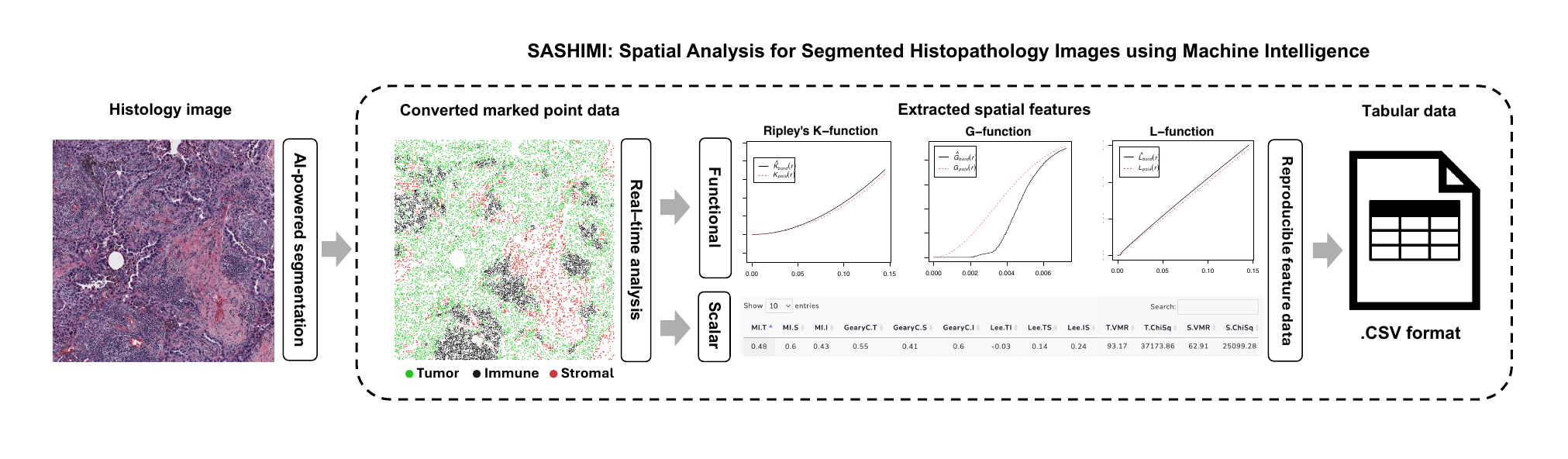}
}
\caption{SASHIMI workflow}
\label{fig:workflow}
\end{figure}

The overall analytical workflow of the SASHIMI is illustrated in Figure \ref{fig:workflow}, which outlines the end-to-end computational process from raw histopathology images to quantitative spatial features. SASHIMI operates through three intuitive steps: (1) uploading marked point pattern data; (2) selecting desired spatial feature categories; and (3) visualizing and downloading the computed results.

\begin{figure}[htbp]
\centering
\includegraphics[width=1.1\textwidth]{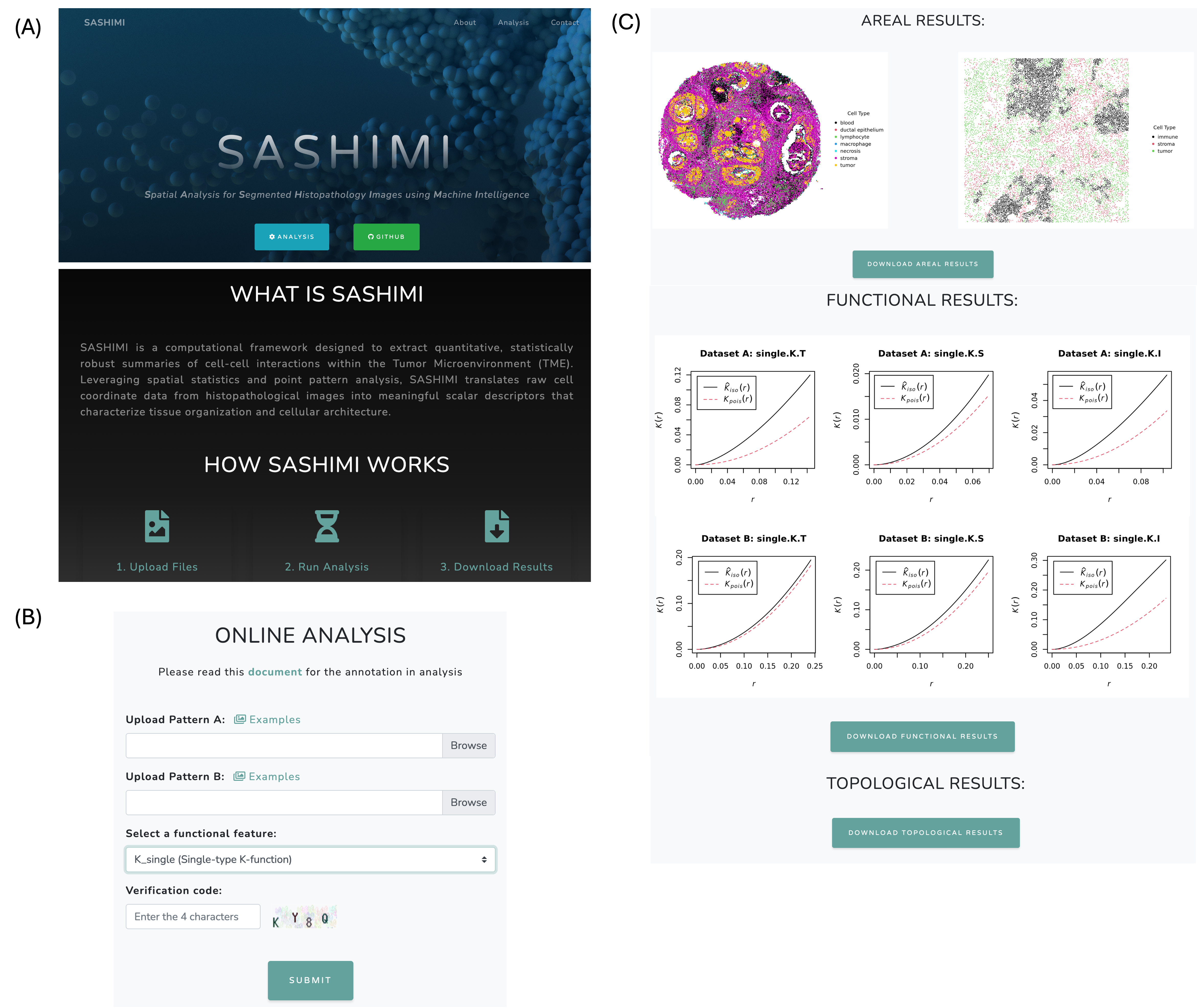}
\caption{User interface of the SASHIMI web application.
(A) Home and overview interface introducing the framework and navigation structure; 
(B) Online analysis page for uploading point-pattern data and selecting spatial features; 
(C) Output page displaying areal-level maps and functional results with download options.
}
\label{fig:sashimi_ui}
\end{figure}

The corresponding web interface of SASHIMI is shown in Figure \ref{fig:sashimi_ui}, which demonstrates the user-facing implementation of this pipeline.
Panel (A) shows the home and overview interface, where users are introduced to the purpose and functionality of SASHIMI through concise descriptions and navigation links.
Panel (B) displays the online analysis interface, where users can upload one or two point-pattern files (e.g., different cell types or tissue regions), specify functional statistics such as the $K$-function, and submit the job for computation.
Panel (C) illustrates the output interface, which presents both areal-level spatial maps and functional summary plots (e.g., $K$-, $G$-, and $L$-functions), along with downloadable CSV tables containing all computed indices.

Together, these components provide a seamless and reproducible environment for quantitative spatial analysis of AI-segmented histopathology images. SASHIMI is freely accessible as a web application at \url{https://lce.biohpc.swmed.edu/sashimi/}, with full source code and documentation available at \url{https://github.com/hangookminsokchon/SASHIMI}.

\section{Case Studies}\label{sec-case}

To illustrate the use of SASHIMI as an exploratory spatial analysis and feature extraction platform,
we analyzed two real-world datasets: lung cancer pathology images from NSCLC patients in the NLST and oral tissue pathology images from OPMD patients in the EPOC trial at The University of Texas MD Anderson Cancer Center.


Rather than performing a comprehensive evaluation or comparative assessment of spatial features, these case studies are intended to demonstrate how SASHIMI facilitates systematic extraction, visualization, and organization of diverse spatial descriptors from AI-segmented histopathology images.
To provide a concrete illustration of downstream analysis,
functional Cox proportional hazards (FCoxPH) models were fitted using the \texttt{survival} package in \texttt{R}. 
Patient ID was used as a clustering variable to account for within-patient correlations arising from multiple image regions. Each spatial feature was modeled individually, with patient age and cancer stage included as additional scalar covariates. Statistical significance was assessed across multiple thresholds 
in an exploratory manner, with the goal of highlighting representative spatial patterns that may warrant further investigation rather than establishing definitive prognostic biomarkers.
To characterize distance-dependent spatial dynamics, we applied functional principal component analysis (FPCA) to the computed functional descriptors. Spatial functions were decomposed using FPCA based on the Karhunen–Loève expansion~\citep{kong2018flcrm}, and the first two principal components, which captured approximately 90\% of the total variance, were retained. The resulting FPCA scores were then used as covariates in the survival analysis models, along with additional scalar covariates. Additional methodological details are provided in the supplementary materials.

\subsection{Case Study I: Lung Cancer Pathology Images}\label{expt:nlst}

The NLST dataset comprised 345 tumor images from 188 patients, acquired from the National Lung Screening Trial (NLST) conducted by the National Cancer Institute. All images were formalin-fixed paraffin-embedded (FFPE) slides captured at $40\times$ magnification. Following segmentation, $1585$ image regions were obtained, each containing approximately $12,000$ individual cell points. For model fitting, $1578$ images were used after excluding seven images with missing cell-type annotations.

Several spatial features demonstrated statistically significant associations with survival outcomes. Table~\ref{tab:assoc_nlst} summarizes the main predictors identified at different $p$-value thresholds, spanning classical spatial statistics, topological descriptors, and areal indices. The complete set of results is provided in Supplementary Table 2.

At the most stringent level ($p < 0.05$), two topological features and one spatial summary statistic feature were identified. Specifically, \texttt{witness\_tumor\_stromal\_h0\_lifetime\_max} and \texttt{witness\_tumor\_stromal\_h0\_death\_std} emerged as highly significant, underscoring the relevance of connected-component persistence in tumor-stromal organization. In addition, the \texttt{PCF.CROSS.T2I.PC1} captured pairwise correlation dynamics between tumor and immune cells.

When the threshold was relaxed to $p < 0.10$, additional predictors became significant, including \texttt{MK.CONN.S2I.PC2} and \texttt{L.CROSS.T2I.PC1}, both reflecting Cross-type clustering dynamics, as well as two topological $H_1$ features, \texttt{witness\_immune\_stromal\_h1\_birth\_std} and \texttt{witness\_tumor\_immune\_h1\_death\_mean} representing the variability and persistence of loops across immune-stromal and tumor-immune interactions.

At $p < 0.15$, further single-type and scalar indices (\texttt{J.REP.I.PC1}, \texttt{G.CROSS.T2I.PC1}, \texttt{MH.IS}) were identified, along with an additional Cross-type $K$-function (\texttt{K.CROSS.T2S.PC2}). By $p < 0.20$, a broader set of spatial and topological descriptors reached significance, including additional Cross-type $K$-, $J$-, $F$-, and $G$-functions.

Collectively, these findings demonstrate that conventional spatial statistics and topological descriptors provide complementary insights into the spatial organization of the tumor microenvironment. Notably, Cross-type interactions consistently emerged as the most prognostic features, outperforming single-type or type-agnostic measures. This pattern underscores the importance of explicitly incorporating cell-type information when characterizing spatial proximity to tumor cells, aligning with established biological principles regarding the functional role of intercellular interactions in cancer progression ~\citep{hanahan2011hallmarks}. In particular, immune–tumor and immune–stromal interactions exhibited the strongest statistical associations with survival outcomes, whereas stromal–tumor relationships showed weak or negligible correlations with patient prognosis.

When examined by feature class, topological descriptors derived from the witness complex and pair correlation functions achieved statistical significance at the most stringent threshold. Although both the pair correlation function and the K-function quantifies distance-dependent clustering, the K-function failed to capture meaningful prognostic signals in this setting. This limitation arises from its cumulative formulation, which aggregates the number of neighboring cells within an increasing radius r, thereby emphasizing overall abundance rather than local spatial dynamics. In contrast, the pair correlation function evaluates cell interactions at fixed spatial scales, isolating distance-specific deviations from spatial randomness. This distinction enabled the detection of prognostically relevant immune–tumor spatial relationships, reinforcing the importance of spatial proximity over coarse measures of cellular abundance within the tumor microenvironment~\citep{shen2024spatial, backman2023spatial, yuan2016spatial}.

\begin{table}[!ht]
	\centering
	\footnotesize
	\renewcommand{\arraystretch}{0.5} 
	\caption{Significant spatial and topological features associated with survival outcomes in the NLST dataset.}
	\label{tab:assoc_nlst}
	\begin{tabular}{@{}l l r@{}}  
		\toprule
		\textbf{Threshold} & \textbf{Feature} & \textbf{$p$-value} \\
		\midrule
		\multirow{3}{*}{$p < 0.05$} 
		& \texttt{witness\_tumor\_stromal\_h0\_lifetime\_max} & 0.005 \\
		& \texttt{witness\_tumor\_stromal\_h0\_death\_std} & 0.018 \\
		& \texttt{PCF.CROSS.T2I.PC1} & 0.028 \\
		\midrule
		\multirow{4}{*}{$p < 0.10$}
		& \texttt{witness\_immune\_stromal\_h1\_birth\_std} & 0.062 \\
		& \texttt{MK.CONN.S2I.PC2} & 0.063 \\
		& \texttt{L.CROSS.T2I.PC1} & 0.065 \\
		& \texttt{witness\_tumor\_immune\_h1\_death\_mean} & 0.098 \\
		\midrule
		\multirow{4}{*}{$p < 0.15$}
		& \texttt{J.REP.I.PC1} & 0.104 \\
		& \texttt{G.CROSS.T2I.PC1} & 0.109 \\
		& \texttt{K.CROSS.T2S.PC2} & 0.143 \\
		& \texttt{MH.IS} & 0.144 \\
		\midrule
		\multirow{5}{*}{$p < 0.20$}
		& \texttt{K.CROSS.S2I.PC1} & 0.155 \\
		& \texttt{J.REP.PC2} & 0.158 \\
		& \texttt{K.CROSS.T2S.PC1} & 0.161 \\
		& \texttt{F.REP.I.PC2} & 0.165 \\
		& \texttt{G.REP.PC2} & 0.191 \\
		\bottomrule
	\end{tabular}
\end{table}

\subsection{Case Study II: Oral Tissue Pathology Images}\label{expt:opmd}
The OPMD dataset consisted of $255$ whole-slide images acquired at $40\times$ magnification from 128 patients with oral premalignant lesions. Using a deep convolutional neural network developed by our team~\citep{wang2019convpath}, nuclei were segmented and classified into four distinct categories: basal, other epithelial, lymphocytes, and stromal cells. For feature computation, only three types—basal, lymphocyte, and stromal cells—were used, excluding the other epithelial category. This resulted in $703$ segmented image regions, each containing approximately $1000$ cell points.

Spatial analysis of the OPMD dataset revealed distinct predictors associated with patient survival. 
Table~\ref{tab:assoc_opmd} summarizes the significant features identified across different $p$-value thresholds. The complete set of results is provided in Supplementary Table 2.

At the most stringent level ($p < 0.05$), \texttt{J.REP.T.PC2} emerged as one of the strongest predictors, representing empty-space dynamics in tumor cell organization. 
Notably, topological features were also significant, including the number of loops created (\texttt{witness\_tumor\_stromal\_h1\_n\_features}) and the number of connected components (\texttt{witness\_tumor\_immune\_h0\_n\_features}), indicating the importance of loop and connected component counts in tumor-stromal and tumor-immune  organization.

When the threshold was relaxed to $p < 0.10$, additional predictors became significant, including \texttt{J.REP.PC2}, \texttt{J.CROSS.T2I.PC2}, and \texttt{J.REP.S.PC2}, along with topological features such as \texttt{witness\_tumor\_stromal\_h0\_death\_std} and \texttt{witness\_tumor\_stromal\_h0\_n\_features}.

Expanding the criterion to $p < 0.15$ revealed further Cross-type spatial descriptors, including \texttt{G.CROSS.T2I.PC2} and \texttt{PCF.CROSS.T2I.PC2}. 
At $p < 0.20$, additional features reached significance, such as \texttt{G.REP.PC2}, \texttt{MK.CONN.T2S.PC2}, and \texttt{PCF.CROSS.S2I.PC2}, together with several similarity indices, including \texttt{Jaccard.IS} and \texttt{Dice.IS}.

Overall, the OPMD dataset exhibited a distinct pattern of significant spatial predictors compared with the NLST cohort, with J-functions and topological feature counts playing a particularly prominent role in survival prediction. In contrast to the NLST dataset, both Cross-type and single-type interactions contributed substantially to prognostic modeling in OPMD. Among these, tumor–stromal and tumor–immune interactions demonstrated the strongest associations with survival outcomes, followed by tumor-only interaction patterns. Notably, several single-type and aggregation-based spatial functions emerged as significant predictors in OPMD, despite showing little or no prognostic relevance in the NLST cohort. In contrast, topological features consistently retained their predictive value across both datasets, underscoring their robustness to cohort-specific variation.

Given the smaller segmented batch size of the OPMD dataset—reflecting inherent differences in lesion size, with approximately 1,000 cell points per image compared to an average of 12,000 in the NLST dataset—the enhanced performance of aggregation-based functions such as the J-function is likely attributable to their sensitivity to local spatial structure. This suggests that, under constrained image sizes, spatial summary functions may preferentially capture fine-scale organization that becomes diluted in larger fields of view. Consequently, these findings highlight the batch-size sensitivity of certain spatial descriptors and emphasize the importance of considering image scale and sampling density when interpreting spatial feature performance across heterogeneous cohorts.

\begin{table}[!ht]
	\centering
	\footnotesize
	\renewcommand{\arraystretch}{0.5}  
	\caption{Significant spatial and topological features associated with survival outcomes in the OPMD dataset.}
	\label{tab:assoc_opmd}
	\begin{tabular}{@{}l l r@{}} 
		\toprule
		\textbf{Threshold} & \textbf{Feature} & \textbf{$p$-value} \\
		\midrule
		\multirow{3}{*}{$p < 0.05$}
		& \texttt{witness\_tumor\_stromal\_h1\_n\_features} & 0.011 \\
		& \texttt{J.REP.T.PC2} & 0.020 \\
		& \texttt{witness\_tumor\_immune\_h0\_n\_features} & 0.023 \\
		\midrule
		\multirow{5}{*}{$p < 0.10$}
		& \texttt{J.REP.PC2} & 0.067 \\
		& \texttt{J.CROSS.T2I.PC2} & 0.071 \\
		& \texttt{J.REP.S.PC2} & 0.083 \\
		& \texttt{witness\_tumor\_stromal\_h0\_death\_std} & 0.090 \\
		& \texttt{witness\_tumor\_stromal\_h0\_n\_features} & 0.099 \\
		\midrule
		\multirow{2}{*}{$p < 0.15$}
		& \texttt{G.CROSS.T2I.PC2} & 0.117 \\
		& \texttt{PCF.CROSS.T2I.PC2} & 0.142 \\
		\midrule
		\multirow{5}{*}{$p < 0.20$}
		& \texttt{PCF.CROSS.S2I.PC2} & 0.165 \\
		& \texttt{Dice.IS} & 0.171 \\
		& \texttt{Jaccard.IS} & 0.172 \\
		& \texttt{MK.CONN.T2S.PC2} & 0.178 \\
		& \texttt{G.REP.PC2} & 0.187 \\
		\bottomrule
	\end{tabular}
\end{table}

\subsection{Discussion}
A key advantage of integrating multiple classes of spatial descriptors within SASHIMI is the ability to characterize complementary aspects of cellular organization across spatial scales. In this section, we aim to provide a detailed overview of the suite of features by discussing their respective capabilities and limitations.

Radial distribution functions, such as the pair correlation function and mark connection function, emphasize local spatial interactions by quantifying deviations from spatial randomness at specific distance ranges. These features are particularly sensitive to pairwise cellular proximity and localized patterns of structure or avoidance, which are frequently considered biologically relevant in tumor–immune interactions. In contrast, other members of the spatial summary statistics family—namely cumulative descriptors such as Ripley’s K-, G-, and L-functions—aggregate spatial information over increasing radii, capturing more global or multi-scale trends in cellular aggregation. While these cumulative summaries provide a broad overview of spatial organization, they may dilute fine-scale interaction signals when cell density is high or when spatial heterogeneity varies across regions or cohorts. When applied to the two real datasets, radial distribution functions demonstrated stronger correlations with survival outcomes in the NLST dataset compared to the OPMD dataset, where the average number of cells per image batch is substantially larger. Conversely, aggregation functions showed stronger correlations in the OPMD dataset and more moderate associations in NLST, suggesting that prognostic signals in OPMD may be driven by local proximity structure of cells rather than overall abundance or composition~\citep{shen2024spatial, backman2023spatial, yuan2016spatial}.

While the inclusion of both distance-specific and cumulative descriptors within SASHIMI allows users to examine spatial organization from complementary local and global perspectives without presupposing a single “optimal” summary statistic, the scope of spatial structure captured by each feature remains dependent on the choice of image batch size.

Topological descriptors derived from persistent homology offer a distinct perspective on spatial organization by capturing connectivity and structural patterns that are invariant to continuous deformations and less sensitive to local fluctuations in cell density. Unlike distance-based summaries, which depend explicitly on spatial scale parameters and image size, topological features summarize the birth, persistence, and disappearance of connected components and loops across multiple spatial resolutions. In the context of tumor tissue architecture, these properties make topological features well suited for capturing higher-order structural organization, such as compartmentalization and multiscale connectivity between tumor, stromal, and immune regions. Their consistent emergence as some of the strongest covariates across datasets in the illustrative case studies suggests that topological summaries may provide robust, complementary information beyond classical proximity-based measures, particularly in heterogeneous or sparsely sampled imaging settings.

Lastly, the areal feature family exhibited a moderate level of correlation in the OPMD dataset but weak or no meaningful association in the NLST dataset. Similar to aggregation-based functions, areal features computed on a fixed grid resolution limit the precise capture of fine local structure within the tumor microenvironment, as they primarily reflect compositional similarity and regional overlap based on cell abundance.

Another recurring pattern observed across the case studies is the frequent prominence of Cross-type spatial descriptors relative to single-type summaries. Cross-type features explicitly quantify spatial relationships between distinct cell populations, such as tumor–immune or tumor–stromal interactions, thereby encoding biologically meaningful aspects of the tumor microenvironment that cannot be inferred from single-population distributions alone, further recognizing the strong prognostic relevance of cell-type-specific spatial organization.

By systematically organizing both single-type and Cross-type spatial descriptors, SASHIMI enables users to explore these interaction-driven patterns in a unified and interpretable manner, facilitating hypothesis generation regarding the spatial mechanisms underlying disease progression. Importantly, SASHIMI is not designed to rank or prioritize spatial features based on predictive performance. Instead, its primary goal is to provide a standardized and accessible interface for computing, visualizing, and organizing diverse spatial descriptors, allowing downstream statistical modeling and domain-specific interpretation to be conducted as appropriate for individual study designs. This design choice reflects the exploratory and hypothesis-generating role of spatial feature analysis in computational pathology, particularly in settings where spatial heterogeneity and sampling variability are inherent. Complete list of visual examples by significant feature can be found in the supplementary document.

\subsection{Sensitivity of Exploratory Findings to Significance Thresholds}

In the NLST dataset, $16$ spatial features reached significance at $p < 0.20$, with $11$ features remaining significant at $p < 0.15$, $7$ features at $p < 0.10$, and $3$ features at $p < 0.05$. Similarly, the OPMD dataset demonstrated $15$ significant features at $p < 0.20$, $10$ features at $p < 0.15$, $8$ features at $p < 0.10$, and $3$ features achieving significance at $p < 0.05$.
This monotonic reduction in the number of associated features across increasingly stringent thresholds reflects the sensitivity of exploratory spatial findings to threshold choice, rather than providing evidence of optimal cutoff values.
These observations indicate that spatial descriptors may exhibit moderate and context-dependent associations with clinical outcomes, which is consistent with the inherent variability of cellular spatial arrangements in tumor tissue architecture and heterogeneity across disease sites.

The use of multiple significance thresholds is justified by several factors inherent to spatial point pattern analysis. First, the resolution and density of marked point pattern data can influence the sensitivity of spatial statistics to detect biologically relevant structures. Second, the complex multi-scale nature of tumor spatial organization may yield moderate rather than strong statistical associations. 
Third, given the exploratory role of comprehensive spatial feature analysis, considering a range of nominal significance thresholds allows potentially informative spatial patterns to be identified for further investigation in larger cohorts or independent validation studies, rather than being prematurely excluded by a single stringent cutoff.

\section{Conclusion}\label{sec-conclusion}
The spatial organization of cells within tumor tissue carries critical information about disease progression and patient outcomes, yet accessible computational tools for rigorous quantification of these spatial patterns have remained limited. We developed SASHIMI, a web-based framework that integrates diverse spatial statistics, topological features, and areal similarity indices to enable systematic extraction and analysis of spatial biomarkers from AI-segmented histopathology images.

SASHIMI addresses key limitations in existing approaches by providing a comprehensive suite of $27$ spatial features derived from single-type, Cross-type interactions, and topological characteristics. To our knowledge, this is the first publicly available platform that combines this breadth of spatial analytical methods with real-time computation and interactive visualization in an accessible web interface. The framework extends beyond exploratory visualization by enabling direct integration of extracted features into downstream statistical and machine learning workflows, facilitating spatial biomarker discovery for researchers without specialized computational expertise.

While designed for computational pathology applications, the framework's foundation in marked point pattern analysis makes it applicable to any spatial context involving point pattern data with categorical labels. Future developments will focus on enhancing computational scalability through parallel processing architectures and incorporating automated nuclei detection capabilities to enable direct analysis of raw histopathology slides. By bridging rigorous spatial statistics with accessible biomedical applications, SASHIMI provides a systematic approach to quantifying spatial inter-play between cells in tumor tissue organization and investigating how spatial architecture influences disease progression and clinical outcomes.


\section{Funding}\label{sec-funding}
This work was partially supported by the National Institutes of Health grants R15 CA274241 (C.M.), R01GM141519 (G.X. and Q.L.), R01GM140012 (G.X.), R01GM115473 (G.X.), U01CA249245 (G.X.), the Cancer Prevention and Research Institute of Texas grant CPRIT RP230330 (G.X.), and the National Science Foundation grants DMS-2113674 (Q.L.) and DMS-2210912 (Q.L.)



\section{Data Availability Statement}\label{data-availability-statement}
\noindent \textbf{Case study I.} 
The H\&E-stained pathology images and patient clinical information analyzed in this study are available through the National Lung Screening Trial (NLST) data portal at \url{https://cdas.cancer.gov/learn/nlst/home/}. 
Access to the raw images requires submission of a data request through the Cancer Data Access System (CDAS). 
The AI-segmented images generated in this study are available from the corresponding author upon reasonable request.

\noindent \textbf{Case study II.} 
The OPMD dataset analyzed in this study is available in the Zenodo repository under restricted access at \url{https://doi.org/10.5281/ZENODO.10664056}.
This dataset is described in~\citep{piyarathne2024comprehensive}.
The nuclei segmentation and classification results generated in this study using ConvPath~\citep{wang2019convpath} are available from the corresponding author upon reasonable request.


\phantomsection\label{supplementary-material}
\bigskip

\begin{center}


\end{center}



   \bibliography{bibliography_edit.bib}

\end{document}